\documentclass[a4paper,aps,prl,twocolumn,showpacs,amsmath,amssymb]{revtex4}
\usepackage{amsmath}
\usepackage{graphicx}
\begin{document}

\title{Light-Matter Quantum Interface}

\author{K. Hammerer$^1$, K. M$\o$lmer$^{2,3}$, E.S. Polzik$^{2,4}$, J.I. Cirac$^1$}

\affiliation{ $^1$Max-Planck--Institut f\"ur Quantenoptik,
Hans-Kopfermann-Strasse, D-85748 Garching, Germany \\
$^2$ QUANTOP, Danish Research Foundation Center for Quantum Optics\\
$^3$ Department of Physics and Astronomy, University of Aarhus, DK 8000
Aarhus C, Denmark \\
$^4$ Niels Bohr Institute, DK 2100 Copenhagen, Denmark}

\begin{abstract}
We propose a quantum interface  which applies multiple passes of a
pulse of light through an atomic sample with phase/polarization
rotations in between the passes. Our proposal does not require
nonclassical light input or measurements on the system, and it
predicts rapidly growing entanglement of light and atoms from just
coherent inputs. The proposed interface makes it possible to
achieve a number of tasks within quantum information processing
including teleportation between light and atoms, quantum memory
for light and squeezing of atomic and light variables.
\end{abstract}

\pacs{03.67.Mn, 32.80.Qk}

\maketitle


Quantum networks require an efficient quantum interface between
light, which is a natural long distance carrier of quantum
information, and atoms, that make a better storage and processing
medium. The power of such a device will be intimately connected to
its capability of creating high degrees of entanglement in a
controlled way, since entanglement represents an all-purpose
resource to create conditional dynamics.

\begin{figure}[t]
\includegraphics[width=6cm]{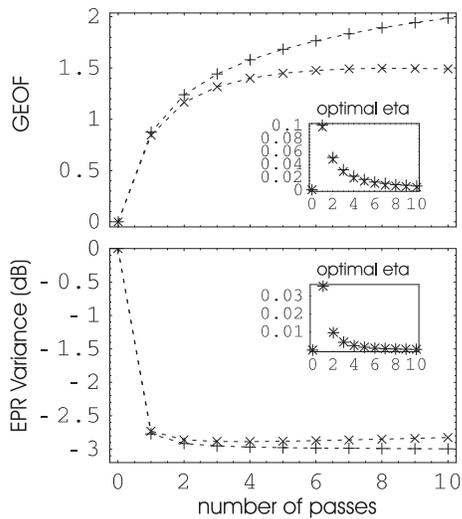}
\caption{GEOF and EPR variance vs. number of passes: For given $n$
both quantities are maximized with respect to $\eta$ and $\zeta$.
The optimal values for $\eta$ are shown in the inserts. It is
always best to have $\zeta = r$ corresponding to
$\epsilon\ll\eta$. $+'$es refer to the case $r=0$, $\times'$es to
$r=2\%$. The optical density is $\alpha_0 = 25$.
\label{unswitched}}
\end{figure}

Numerous theoretical and experimental works (\cite{KMP} and
references therein) center around the effect of a Kerr interaction
between light and atomic ensembles to produce entanglement between
continuous light-atom variables. This interspecies entanglement
can in turn be converted to atomic spin correlations in form of
spin squeezed or spin entangled states between two atomic samples
\cite{JKP} by means of a projection measurement on light. If
spontaneous emission is neglected, the degree of squeezing is of
the order of $\Delta=1/(1+\kappa)$ where $\kappa$ is the effective
coupling strength between light and atoms. Thus, it seems that one
can, in principle, produce in this way unlimited atomic squeezing.
There are however serious limitations on both, the amount of light
atom entanglement as well as the degree of squeezing, which can be
achieved from a Kerr interaction. In fact one can express
$\kappa=\alpha_0\eta$ where $\alpha_0$ is the sample's optical
(column) density and $\eta$ is the spontaneous emission
probability. Thus, decoherence due to spontaneous emission cannot
be neglected and a crude estimate for $\eta\ll 1$ leads to an
additional contribution, so that
$\Delta=1/(1+\alpha_0\eta)+2\eta$ ($\Delta=1$ corresponds to a
coherent state). Apparently this expression has a minimum
$\Delta_{min}=2\sqrt{2/\alpha_0}$ corresponding to an optimal
spontaneous decay probability $\eta_0=1/\sqrt{2\alpha_0}$. In
realistic systems the optical density is often limited to values
in the range between 1-100, which is true for atomic vapors (ref.
[7]), as well as for cold and trapped atoms. For the optical
density of, say, 25, the above estimates lead to the limit on
squeezing of the order of $\Delta_{min}\approx 0.5$ (3 dB of noise
reduction) with a single pass QND measurement. The same
consideration limits also the amount of light atom entanglement
present before the measurement (see Fig.\ref{unswitched}).

But there is still another peculiarity of the Kerr interaction,
which limits its performance in creating entanglement: Due to its
QND character this interaction conserves certain degrees of
freedom, which is reflected in a strict limit on the amount of
achievable EPR-type squeezing to $\Delta EPR=0.5$. Thus, even for
an arbitrarily high optical density, the state which originates
from a Kerr interaction is never close to a maximally entangled
EPR state corresponding to $\Delta EPR\rightarrow 0$.

In this Letter we propose experimentally feasible techniques which
allow to overcome these limitations and in fact provide an
exponential growth in the amount of entanglement and squeezing.
Two and three pass protocols have already been proposed in
\cite{F,K,KMP}. Taking spontaneous decay into account, we show
that passing one and the same pulse of light $n$ times through an
atomic ensemble produces an effective optical density of
$n\alpha_0$ while the effect of accumulated spontaneous emission
noise can be balanced by tuning $\eta$ to its optimal value for a
given number of steps $n$. Hence, although the coupling strength
in a Kerr interaction is directly proportional to the probability
of spontaneous decay, this does not pose a fundamental limit on
the generation of entanglement or squeezing.

Based on this result we show furthermore that this system provides
the realistic possibility to implement a quantum information
algorithm proposed in \cite{KHGC} for generating entanglement and
squeezing at optimal rates in pure Gaussian continuous variable
states. This algorithm is based on the idea to intersperse
interactions with local operations and to optimize these local
operations such as to maximally increase the quantity of interest.
In the system under consideration these optimal local operations
can be effected simply by $\lambda/4$ plates and mirrors changing
the polarization and direction of light propagation in between the
passes of the atomic sample. We determine for realistic
experimental parameters the optimal spontaneous decay probability
for a given number of steps and show thereby that even in the
presence of losses the growth of entanglement is still
significantly enhanced. In particular, one can in this way
engineer a state which is close to an EPR state.

Finally we also suggest a way to convert the entanglement
unconditionally into squeezing of the atoms without the use of
homodyne detection of light. This method which relies on a certain
choice of polarization rotations is as powerful as the QND
measurement and it yields in addition a squeezed optical output.


The system we are considering is the same as in
\cite{JKP},\cite{KBM},\cite{DCZP}. It is shown there that the
interaction between an off-resonant laser pulse and an ensemble of
atoms with total angular momentum equal to $\hbar/2$ is
appropriately described by a Kerr-interaction Hamiltonian $H
\propto J_z S_z$ with collective atomic spin operator $J_z={1\over
2} \sum_i \sigma_z^{(i)}$ and Stokes operator $S_z=(a_R^\dagger
a_R-a_L^\dag a_L)/2$. Here $\sigma_z^{(i)}$ is the Pauli spin
operator along $z$ for the $i-$th atom and $a_R \; (a_L)$ is the
photon annihilation operator for right (left) circularly polarized
photons. In the limit of large polarization along the
$x-$direction one can treat $J_x$ and $S_x$ as classical variables
and represent the orthogonal directions by canonical operators
$x_{at}=J_y/\sqrt{\langle J_x\rangle},\; p_{at}=J_z/\sqrt{\langle
J_x\rangle},\; x_{ph}=S_y/\sqrt{\langle S_x\rangle},\;
p_{ph}=S_z/\sqrt{\langle S_x\rangle}$. Both the atomic coherent
spin state and the polarized laser pulse correspond in this limit
to the ground states of the harmonic oscillators associated with
these operators. The interaction Hamiltonian in the oscillator
formulation  becomes $H \propto \sqrt{\langle J_x\rangle\langle
S_x\rangle} p_{at} p_{ph}$. Since the initial states are Gaussian
distributions over the phase space and the Hamiltonian is bilinear
in the canonical operators and therefore conserves the Gaussian
character of the state it is appropriate to express the dynamics
in the Schr\"odinger picture in terms of a displacement vector
$\vec{d}=\textrm{tr}\{\vec{R}\rho\}$ and a covariance matrix
$\gamma_{i,j}=\textrm{tr}\{\rho[(\vec{R}_{i}-\vec{d}_i),(\vec{R}_{j}-\vec{d}_j)]_+
\} \quad (i,\:j = 1,\ldots,4)$ where $\vec{R} =
(x_{at},p_{at},x_{ph},p_{ph})$ and $[.,.]_+$ denotes the
anticommutator. For the given initial states and within the above
approximation we have $\vec{d}=\vec{0}$ for all times. Thus, all
information about the compound system can be extracted from its
covariance  matrix.


The state after a single pass of a pulse of light through the
atomic ensemble is described in terms of input-output relations as
\begin{equation}\label{InOutSingle}
\gamma_{out} = \bar{D}(\eta,\epsilon) S(\kappa) \gamma_{in}
S(\kappa)^T \bar{D}(\eta,\epsilon) + D(\eta,\epsilon)
\gamma_{noise}
\end{equation}
where the scattering matrix
\begin{equation}\label{ScatteringMatrix}
S(\kappa) = \left(
\begin{array}{llll}
1 & 0 & 0 & \kappa \\
0 & 1 & 0 & 0 \\
0 & \kappa & 1 & 0 \\
0 & 0 & 0 & 1 \\
\end{array}
\right)
\end{equation}
and $D(\eta,\epsilon) =
\textrm{diag}(\eta,\eta,\epsilon,\epsilon), \;
\bar{D}(\eta,\epsilon) = \sqrt{\textbf{1}-D(\eta,\epsilon)}, \;
\gamma_{noise} = \textrm{diag}(2,2,1,1)$. The output state is a
weighted sum of a coherent contribution and a noise component
$\gamma_{noise}$ whose form is due to the fact that the field
decay is accompanied by  a vacuum noise contribution and the
atomic decay both contributes to noise due to the breaking of
correlations among the atoms and due to the atoms once decayed
being still present in the sample, explaining the factor of 2 in
the atomic component of $\gamma_{noise}$. Apart from this
correction Eq. (\ref{InOutSingle}) is equivalent to the result
derived in \cite{DCZP}. In principle, the noise introduced in
atoms increases with the decay of the mean polarization, but this
effect is negligible for the example presented (see \cite{MM} for
a refined model for this interaction using the same formalism).

The coupling constant is given by $\kappa = 2\sqrt{\langle
J_x\rangle\langle S_x\rangle}\sigma\Gamma/A\Delta $, the atomic
depumping $\eta = N_{ph} \sigma\Gamma^2/A\Delta^2$ and the
photonic absorption rate $\epsilon =
N_{at}\sigma\Gamma^2/A\Delta^2$ where $\sigma$ is the cross
section on resonance for the probed transition, $\Gamma$ is the
corresponding spontaneous decay rate, $\Delta$ the detuning from
resonance and $A$ the cross section of the atomic ensemble
illuminated by the pulse. Equation (\ref{InOutSingle}) is valid
for small atomic dephasing and low photon absorption corresponding
to $\eta,\epsilon \ll 1$.

A central quantity in this system is the optical density on
resonance $\alpha_0 = N_{at}\sigma/A$ which gives the probability
for a single photon to get elastically scattered and can be
related to the other parameters as $\epsilon = \alpha_0
(\Gamma/\Delta)^2$ and $\kappa^2 = \eta \alpha_0$ where we used
that initially $\langle J_x\rangle = N_{at}/2$ and $\langle
S_x\rangle = N_{ph}/2$. There is an apparent tradeoff between
having a large coupling and at the same time low atomic depumping.
For a given optical density one can treat $\epsilon$ and $\eta$ as
independent parameters tailoring the first by means of the
detuning and the last by means of $N_{ph}$, and there are always
optimal values for $\epsilon$ and $\eta$ which maximize the
achievable squeezing or entanglement.

We are here especially interested in three quantities
characterizing the quantum properties of the state generated: (a)
the Gaussian Entanglement of Formation (GEOF) \cite{WGKWC}, the
only available physical Entanglement measure for mixed Gaussian
bipartite states, (b) the closely related \cite{GWKWC} EPR
uncertainty of the combined atom+field system, which indicates how
close the state is to a maximally entangled EPR state, given for
the present states by $\Delta
EPR=\frac{1}{2}[\Delta^{2}(x_{at}-p_{ph})+\Delta^{2}(p_{at}-x_{ph})]$,
and finally (c) the atomic (and light) squeezing achievable either
by a QND measurement (homodyne detection of light) or by means of
a particular disentangling operation at the end of the multi pass
protocol.


The state created after several passes can be calculated by
iterating the map defined by equation (\ref{InOutSingle}). Note
however that the coupling strength depends on the polarizations
along $x$ and that these classical variables will decay from pass
to pass as $\langle J_x\rangle_{out} = (1-\eta) \langle
J_x\rangle_{in},\; \langle S_x\rangle_{out} = (1-\epsilon) \langle
S_x\rangle_{in}$. For the $n$-th step the remaining coupling
strength is hence reduced $\kappa_n =
[(1-\eta)(1-\epsilon)]^{n/2}\kappa$. Reflection losses can be
taken into account by replacing $\epsilon$ by $\zeta = \epsilon +
r$ where $r$ is the overall reflectivity of mirrors, cell etc.
Equation (\ref{InOutSingle}) provides then readily a recursion
relation
\begin{equation}\label{InOutMultiple}
\gamma_{n} = \bar{D}(\eta,\zeta) S(\kappa_n) \gamma_{n-1}
S(\kappa_n)^T \bar{D}(\eta,\zeta) + D(\eta,\zeta) \gamma_{noise}
\end{equation}
for the state after $n$ passes which can be solved exactly.

The effect of $n$ consecutive passes is comparable to that of a
single pass performed with an $n$ times increased optical density.
This is clear from the meaning of $\alpha_0$ and becomes manifest
in the group property $S(\kappa)S(\lambda) = S(\kappa+\lambda)$ of
the scattering matrix (\ref{ScatteringMatrix}). This indicates
that the strategy of multiple passes is especially interesting for
low optical densities. The dependence of the GEOF and the EPR
variance on the number of passes is shown in figure
\ref{unswitched}. In general it can be shown under the assumption
of vanishing reflection losses $(r = 0)$ that for given optical
density and number of steps $n$ there exist optimal choices for
$\eta$ and $\epsilon$ such that, taking formally $n \rightarrow
\infty$, the GEOF tends to infinity. The EPR-variance is limited
by $0.5$, or 3 dB of squeezing, which is also evident in figure
\ref{unswitched}.

\begin{figure}[t]
\includegraphics[width=6cm]{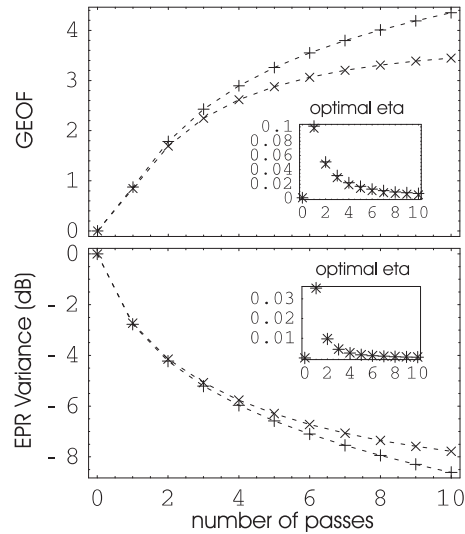}
\caption{GEOF and EPR variance vs. number of passes including
polarization rotations: $+'$es refer to the case $r=0$,
$\times'$es to $r=2\%$. Optical density $\alpha_0 = 25$.
\label{switched}}
\end{figure}

The multipass scheme is capable of improving these features
significantly. In particular, applying a unitary operation and its
adjoint before and after an interaction changes effectively the
type of interaction due to the identity $U^\dagger \exp(-iH)U =
\exp(-i U^\dagger H U)$. The transformations which are easy to
perform in this system are polarization rotations which change the
quadratures as $x \rightarrow \cos \phi\, x +\sin \phi\, p,\; p
\rightarrow \cos \phi\, p - \sin \phi\, x$. In \cite{KHGC} it was
shown in a pure state analysis that entanglement and squeezing is
created at a maximal rate if one switches from $H \propto
p_{at}p_{ph}$ to an interaction $H \propto - x_{at}x_{ph}$ in
every second step. The effect of the switching becomes clear if
one approximates
$\exp(ix_{at}x_{ph}\kappa)\exp(-ip_{at}p_{ph}\kappa) \simeq
\exp[-i(p_{at}p_{ph}-x_{at}x_{ph})\kappa+o(\kappa^2)]$. To first
order this interaction creates a two-mode squeezed state. In
particular the growth is linear in $n$ and thus provides an
exponential improvement as compared to the scheme without
switching. The final state after $n$ passes follows from equation
(\ref{InOutMultiple}) by taking the scattering matrix to be
$S(\kappa)^T$ - corresponding to an interaction $H \propto -
x_{at}x_{ph}$ - in every second step. Figure \ref{switched} shows
how the quantities of interest develop. In comparison with the
unswitched case, the GEOF is roughly doubled and the EPR squeezing
is no longer limited to 3 dB. In the limit of $n \rightarrow
\infty$ the resulting state approximates a maximally entangled EPR
state which can be used as a resource for continuous variable
teleportation. This provides an attractive possibility to
establish a quantum memory for light since an unknown quantum
state of light can be teleported onto the atoms by performing a
joint measurement on the unknown input state and the optical
component  of the EPR state.

\begin{figure}[t]
\includegraphics[width=6cm]{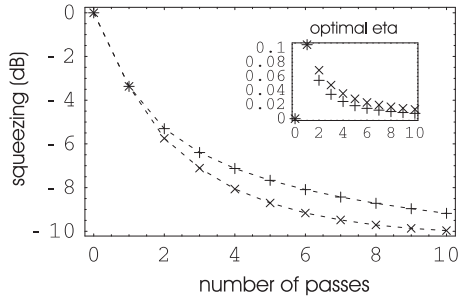}
\caption{Atomic squeezing after homodyne detection of light:
Unswitched scheme (QND measurement) $"+"$ and switched scheme
$"\times"$. $\zeta = r = 2\%,\;\alpha_0 =
25$.\label{QNDSqueezing}}
\end{figure}

After multiple passes (with or without switching of polarizations)
neither light nor atomic quadratures are squeezed separately. In
order to obtain such local squeezing an additional operation has
to be carried out. One possibility is to perform a destructive
homodyne detection of light, which - in the unswitched scheme -
amounts to a QND measurement of the atomic $p$-quadrature and
yields a squeezed state of atoms while the light is lost.
Performing the same measurement on one half of an EPR state - as
it arises in the switched scheme - also leaves the other system in
a squeezed state. Figure \ref{QNDSqueezing} displays the atomic
squeezing after a homodyne detection of light for both schemes.
The switching provides a small advantage even though the actual
interaction has lost its QND character. The tradeoff between
squeezing and spontaneous emission noise has also been discussed
in \cite{AL} for a different type of interaction.

\begin{figure}[t]
\includegraphics[width=6cm]{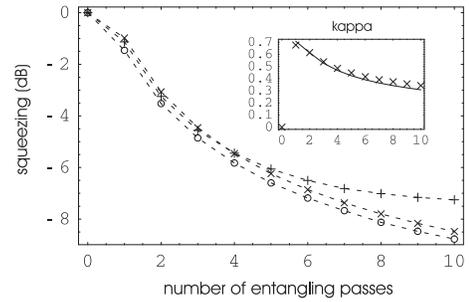}
\caption{Squeezing of light $("+")$ and atomic $("\times")$
quadrature after $n$ entangling and a single disentangling step.
Result from a comparable QND measurement on the atomic system
$"\circ"$. $\zeta = r = 2\%,\;\alpha_0 = 25$. Insert: Optimal
value for coupling $\kappa_{opt} = \alpha_0\eta_{opt}$ and
theoretical magical value $\kappa_0 = \sqrt{n-1/2}/n$ (solid
line). Atomic depumping $\eta$ decreases with $\kappa$ while light
suffers a constant loss of $2\%$ per pass. Therefore the asymmetry
in squeezing of light an atomic
variables.\label{DisentanglingSqueezing}}
\end{figure}

We now show that provided the coupling strength $\kappa$ can be
tuned to a certain value, it is possible to disentangle the state
created after several passes by an appropriate last passage of the
light pulse through the atomic cloud. The basic mechanism is most
clearly seen on the basis of pure states and for the scheme
without polarization switching, but it can be easily adapted also
for the other case. After $n$ passes the atom and field operators
have evolved in the Heisenberg picture as given by $\vec{R'} =
S(n\kappa) \vec{R}_{in}$ where $\vec{R}$ is defined as above. By
switching to an interaction $H \propto x_{at}x_{ph}$ a single
additional pass then yields a state $\vec{R}_{out} = S(-\kappa)^T
\vec{R'}$ and thus $p_{at}^{out} = p'_{at}-\kappa x'_{ph} =
(1-n\kappa^2)p_{at}^{in}-\kappa x_{ph}^{in}$. For $0 < n\kappa^2
\leq 1$ this last pass reduces the weight factor of $p_{at}$
indicating the possibility of squeezing but at the same time it
feeds light noise into the atomic variance. With initial coherent
states one finds $\langle (p_{at}^{out})^2\rangle =
[(1-n\kappa^2)^2+\kappa^2]/2$. This expression can be minimized
with respect to the value of $\kappa$, and the optimal value
$\kappa_0 = \sqrt{n-1/2}/n$ leads to a squeezing of
$\langle(p_{at}^{out})^2\rangle/\langle(p_{at}^{in})^2\rangle =
1/n-1/4n^2$ in comparison with the value $1/(n+1/2)$ achievable in
a QND measurement with the same coupling strength $\kappa_0$. For
large $n$ the difference between these two expressions is
negligible. An important aspect of the decoupling scheme is that
it is not conditioned on a measurement result, and as a side
benefit, light is actually simultaneously squeezed. Figure
\ref{DisentanglingSqueezing} shows the squeezing of both light and
atomic quadratures after such a disentangling step as well as the
result of a comparable QND measurement with identical coupling
$\kappa$.

The experimental feasibility of the proposal is illustrated with
the following example. Consider an ensemble of cold 87Rb atoms
with two ground magnetic states, $F = 1,\, m_F = \pm 1$, forming
the atomic two-level spin system. The light is coupled to these
states via D1 transition (HWHM natural linewidth 2.5 MHz).
Assuming a cylindrical atomic sample with the diameter 100 microns
and the length of 500 microns containing $2\times10^6$ atoms
corresponding to a typical dipole trap density of $5\times10^{11}
cm^{-3}$, a resonant optical density of 25 can be achieved with
the atomic dipole crossection $\sigma=10^{-9}cm^2$. To meet the
optimal condition of light absorption being much less than the
spontaneous emission probability, $\epsilon\ll\eta$, we choose the
light detuning greater than 100 MHz. Then $\epsilon$ is reduced to
less than $1.5\times10^{-3}$. Since $\eta/\epsilon=N_{ph}/N_{at}$,
we can now adjust the optimal value for $\eta$ found from
theoretical graphs in Fig. 1-4 by choosing the optimal number of
photons per pulse. For $\eta$ in the range of $0.01 - 0.1$ the
optimal photon number per pulse is $10^7-10^8$. This number of
photons is close to optimal for shot noise limited balanced
detection. In order to fit the experiment on a table top the
physical length of light pulses should not exceed a few meters,
since the "tail" of the pulse should clear through the sample
before its "head" enters the sample in the next pass. A $3 m$
pulse length corresponds to about 30 MHz Fourier limited bandwidth
which fits well with the detuning somewhat greater than 100 MHz.
Switching of the interaction from X-type to P-type for light can
be achieved simply by passing the light through a $\lambda/4$
plate in between the passes. For atoms this switching can be
achieved by changing the propagation direction of light by 90
degrees.


In summary, we have proposed  a quantum interface between light
and atoms capable of performing valuable tasks in quantum
information processing. By means of several interaction steps and
local operations it is possible to efficiently create entangled
and squeezed states. In particular one can engineer an EPR state
which can act as a resource for a quantum memory for light.
Furthermore we showed that without performing any measurement our
multipass scheme allows to create at the same time spin squeezed
atoms and quadrature squeezed light.

This work was supported by the EU IST project RESQ, the EU grant
QUICOV and the Kompetenznetzwerk Quanteninformationsverarbeitung
der Bayerischen Staatsregierung.

\end{document}